\documentclass[intlimits,twoside,a4paper]{article}
\usepackage{wrapfig}
\usepackage{graphicx}
\usepackage{amssymb,amsmath}
\usepackage[T2A]{fontenc}
\usepackage[cp1251]{inputenc}

\usepackage[eqsecnum]{cmpj2}

\newcommand{\be}{\begin{equation}}
\newcommand{\ee}{\end{equation}}
\newcommand{\bea}{\begin{eqnarray}}
\newcommand{\eea}{\end{eqnarray}}

\issue{2014}{17}{4}{43702}
\doinumber{10.5488/CMP.17.43702}

\title[Thermodynamic and dynamic dielectric properties]%
{Thermodynamic and dynamic dielectric properties of one-dimensional hydrogen bonded ferroelectric of PbHPO$_4$-type}

\author[I.R. Zachek \textsl{et al.}]{I.R. Zachek\refaddr{label2}, R.R. Levitskii\refaddr{label1},
  Ya. Shchur\refaddr{label1}, O.B. Bilenka\refaddr{label2}}
\addresses{
 \addr{label2} National University ``Lviv Polytechnic'', 12~Bandera St., 79013 Lviv, Ukraine
\addr{label1} Institute for Condensed Matter Physics of the National
Academy of Sciences of Ukraine, \\ 1~Svientsitskii St., 79011 Lviv,
Ukraine}

\date{Received June 10, 2014, in final form September 30, 2014}
\authorcopyright{I.R. Zachek, R.R. Levitskii, Ya. Shchur, O.B. Bilenka, 2014}

\begin{document}

\maketitle

\begin{abstract}

Within the modified model of proton ordering of one-dimensional ferroelectric
having hydrogen bonds of PbHPO$_4$-type, their thermodynamic and dynamic
characteristics are studied and calculated taking into account the linear
(by crystal deformations  $\varepsilon_i$ ($i=1,3$) and $\varepsilon_4$) contributions
into the energy of a proton system but without taking into account the tunneling in the
two-particle cluster approximation. There has been obtained a good quantitative description
of the temperature dependence of polarization, static dielectric permittivity, heat capacity
and frequency dependence of dynamic dielectric permittivity at different temperatures for
PbHPO$_4$ and PbHDO$_4$ crystals.


\pacs 77.22.Ch, 77.22.Gm, 77.65.Bn, 77.84.Fa, 77.65.Fs
\end{abstract}

\sloppy

\section{Introduction}

 Ferroelectric properties in PbHPO$_4$ (LHP) and  PbDPO$_4$ (LDP) crystals were disclosed
 in work   \cite{Negran}. At temperature  $T_\textrm{c}^{(\mathrm{H})} = 310$~K in LHP and  $T_\textrm{c}^{(\mathrm{D})} = 452$~K
 in LDP a phase transition of the second order takes phase. These are monoclinic crystals of P2/c
 space group in paraelectric phase  \cite{Negran,Nel}. Ferroelectric phase in LHP is characterized
 by a Pc symmetry having a spontaneous polarization in the direction that forms an angle of $10^\circ$
 with the crystal $a$-axis. The elementary cell of LHP contains two molecules. The parameters of unit
 cell of LHP are as follows $ a = 4.688$~{\AA},  $b = 6.649$~{\AA}, $c = 5.781$~{\AA}, $\beta = 97.11^{\circ}$
 and  of LDP~--- $ a = 4.6855$~{\AA},  $b = 6.6911$~{\AA}, $c = 5.7867$~{\AA}, $\beta = 97.10^{\circ}$.

A characteristic feature of crystal structure of ferroelectric of LHP is the presence of
hydrogen bonds that link the  PO$_4$  tetrahedrons into infinite chains that stretch out
along the  $c$-axis. According to the two possible equilibrium positions, the protons (deuterons)
at these bonds in a paraelectric phase are distributed statistically uniformly, while in
ferroelectric phase there appears a spontaneous asymmetry of population. The data presented
in works   \cite{Blinc,Levr,Kro} testify to the fact that the proton on the hydrogen O-H{\ldots}O
bonds move in two-minimum potentials.  A noticeable change of the phase transition temperature
at deuteration of LHP  \cite{Negran,Nel,Blinc,Levr,Kro,Bre,Smu,Fou,Lop}, as well as of dielectric
 \cite{Negran,Smu} and thermal  \cite{Fou} properties testifies to an important role of a cooperative
behavior of protons in the appearance of ferroelectricity in this type of crystals.

Taking into consideration that the direction of dipole moment of $\nu_{4}(B_{u})$
internal mode is close to the direction of spontaneous polarization  $P_\textrm{s}$ \cite{Shchur,Shchur2},
as well as since this mode shows a considerable temperature softening in a ferroelectric phase,
which is similar to the temperature dependence $P_\textrm{s}(T)$, we may assume that this internal mode
$\nu_{4}$ plays a crucial role in the mechanism of the phase transition in LHP crystal.
The inter-phonon interaction between the low-frequency   $\omega_{s}$  and internal   $\nu_{4}$
modes that have got the same symmetry  $(B_{u})$, close to $T_\textrm{c}$, may result in static deformation
of PO$_{4}$ groups, which causes a static dipole moment of  $\nu_{4}$ mode. It is natural to assume
that the value of $P_\textrm{s}$ is mainly determined by a static dipole moment of PO$_{4}$ groups. The angle
between these moments and the vector of spontaneous polarization is  $2^\circ$. The work \cite{Baj}
estimates the contribution of the polar deformation of PO$_{4}$ groups into LHP to be $\sim 80\% $
of the experimentally observed value of LHP. The phase transition into LHP takes place due to the proton ordering,
phonon anharmonicity and proton-phonon interaction.

Microscopic models of phase transition of LHP were discussed and studied in works  \cite{Car,Blin,Zin,Kor,Cha, Wes,Upa}.
The model discussed in  \cite{Car} corresponds to LDP. Works  \cite{Bli,Blin,Zin, Cha,Upa, Wes} present the LHP model that takes
the tunneling of protons on hydrogen bonds into consideration. Work   \cite{Lev} considers a simple two-sublattice
model of partially deuterated crystal. In should be noted that works   \cite{Lev,Cha} take proton-lattice
interaction into consideration as well. Moreover, work  \cite{Cha} considers the anharmonicity of the lattice
oscillations. Unfortunately, in  \cite{Blin,Cha,Lev} there was used an approximation of mean field while
studying the static and dynamic dielectric permittivity, which is insufficient for LHP. In work  \cite{Zin},
a simple model LHP that takes tunneling into consideration is solved within the approximation of two-particle cluster. Using the Green's function method, the work \cite{Wes} presents temperature dependencies
of relaxation times of LHP and LDP, while work  \cite{Upa} presents dynamic dielectric permittivities. However,
all these works do not put forward the task of describing the corresponding experimental data.

The model of a deformed crystal was beneficially used for a description of dielectric, piezoelectric, elastic,
thermal and dynamic characteristics of quasi-one-dimensional ferroelectric with hydrogen bonds of CsHPO$_4$ type in  \cite{LEV1}.

 This work presents a modified model of proton ordering of hydrogen bonded one-dimensional ferroelectric
 of LHP-type that takes into consideration the linear [by deformations $\varepsilon_i$ (${i}=1,\,3$),
 $\varepsilon_4$] contribution into the energy of the proton system. Within the approximation of two-particle
 cluster, their dielectric, piezoelectric, elastic, thermal and dynamic characteristics are calculated.

\section{Model Hamiltonian of PbHPO$_4$ crystal}

Let us consider a proton subsystem LHP that moves on the O-H{\ldots}O bonds forming zigzag-like
chains along the crystal $c$-axis.

\begin{wrapfigure}{i}{0.5\textwidth}
\begin{center}
\includegraphics[width=0.48\textwidth]{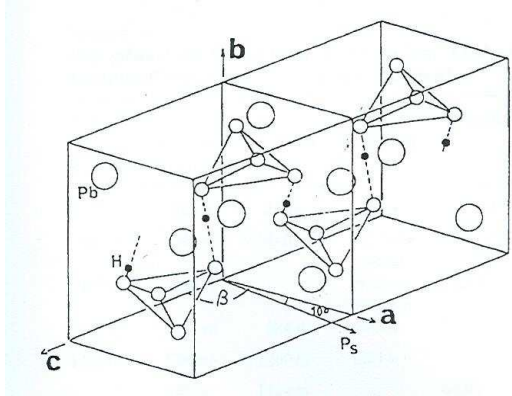}
\end{center}
\vspace{-3mm}
\caption[]{Unit cell of LHP crystal  \cite{Pasq}.} \label{lattice}
\end{wrapfigure}

The unit cell of LHP is formed by the chain that contains two neighboring PO$_4$ tetrahedrons  together with
two short hydrogen bonds that relate to one of them (tetrahedron of `A' type) (see figure \ref{lattice}).
The hydrogen bonds that adjoin the second tetrahedron  (of `B' type) belong to two closest structural elements
of `A' type that surround it.

The Hamiltonian of the proton-ion system of LHP, neglecting the tunneling effects of the protons on the  O-H{\ldots}O bonds,
is as follows  \cite{LEV1,Sta}:
\bea
\label{eq:2.1}
 \hat H &=& NvU_{1 \textrm{seed}} - 2w_1 \sum\limits_{qq'}  \frac{\sigma_{q1}}{2}
\frac{\sigma_{q2}}{2}  \bigl( \delta_{{\bf R}_q{\bf R}_{q'}} + \delta_{{\bf R}_q + {\bf r},{\bf R}_{q'}} \bigr)
- \frac12 \sum\limits_{{qq'},{ff'}}  J_{ff'}(qq') \frac{\sigma_{qf}}{2}\frac{ \sigma_{q'f'}}{2} \nonumber\\
&& + \sum\limits_{kl}\omega_{l}(\vec{k})b_{kl}^{+}b_{kl}+ \frac{1}{\sqrt{N}}\sum\limits_{qf}
\sum\limits_{kl}\sum\limits_{n=1}^2\tau_{lf}(\vec{k})\re^{\ri\vec{k}
\vec{a}_{q}}\left(b_{kl}+b_{-kl}^{+}\right)\left(\frac{\sigma_{qf}}{2}+\frac{\sigma_{qf}}{2}\right)\nonumber\\
&& -\sum\limits_{q}\sum\limits_{kl}\left[\mu_{l}(\vec{k})
{\rm cos}~5^{\circ}E_{1}^{\ast}+\mu_{l}(\vec{k}){\rm cos}~ 2^{\circ}E_s\right]\left(b_{kl}+b_{-kl}^{+}\right)\frac{1}{\sqrt{N}}
\re^{\ri\vec{k}\vec{a_{q}}},
\eea
where $v$ is the volume of unit cell, $N$ is the total number of unit cells, $\sigma_{qf}$
is the $z$-component of pseudo-spin operator that corresponds to the proton located in the $q$-th cell on the
$f$-th bond ($f = 1, \,2$). Eigenvalues of the operator  $\sigma_{qf} = \pm 1$ correspond to two possible positions
of the proton on the hydrogen bond.  Parameters $\mu_p$ and $\mu_{l}(\vec{k})$ are dipole moments that
correspond to the proton on the hydrogen bond and to the dipole-active phonon mode, respectively. $U_{1 \textrm{seed}}$ is
the seed energy that appears in the form of crystal deformations  $\varepsilon_i$, $\varepsilon_4$ and electrical
field  $E_1^{\ast}$ along the crystallographic  axis $a^{\ast}$ which is perpendicular to the plane (b,c),
respectively, and contains elastic, piezoelectric and dielectric parts:
\bea
&& U_{1 \textrm{seed}} = \frac12 \sum\limits_{i,j=1}^3 c_{ij}^{E0} \varepsilon_i\varepsilon_j
+ \sum\limits_{i=1}^3 c_{i4}^{E0} \varepsilon_i\varepsilon_4 + \frac12 c_{44}^{E0}\varepsilon_4^2
- \sum\limits_{i=1}^3 e_{1i}^0E_1\varepsilon_i - e_{14}^0E_1\varepsilon_4  -\frac12 \chi_{11}^{\ast\varepsilon 0}(E_1^{\ast})^2,
\eea
where $c_{ij}^{E0}$, $c_{i4}^{E0}$, $c_{44}^{E0}$, $e_{1i}^0$, $e_{14}^0$, $\chi_{11}^{\ast\varepsilon 0}$
are seed elastic stress, coefficients of piezoelectric strain and dielectric susceptibility of a mechanically
clamped crystal.

The second term in (\ref{eq:2.1}) is the Hamiltonian of short-range interactions between protons.
The first Kronecker's symbol corresponds to the proton interaction in the chain close to the tetrahedron of `A'-type,
the second Kronecker's symbol corresponds to proton interactions near the `B'-type tetrahedron, ${\bf r}$ is
radius vector of the relative position of the proton bond in the unit cell. The quantity $w_1$ that describes
short-range proton interactions within the chain may be expanded into a series with respect to deformations
$\varepsilon_i$, $\varepsilon_4$ restricted to the linear summands  \cite{LEV2}:
\be
\label{eq:2.3}
w_1 = w + \sum\limits_{i=1}^3 \delta_{i}\varepsilon_i + \delta_{4}\varepsilon_4\,.
\ee
The third term describes the effective long-range dipole-dipole interaction between protons within the
chain running along $c$-axis, the fourth summand represents the lattice vibration energy ($b_{kl}^{+}$, $b_{kl}$~--- Bose
operators, ${l}$-index of phonon branch), the fifth summand corresponds to the proton-phonon interactions.
The last term describes the interaction of the lattice with the external electrical field.

Long-range interactions of protons and interactions of protons with the lattice vibrations
are considered within the the mean field approximation. Thus, the Hamiltonian  (\ref{eq:2.1}) looks as follows:
\bea
\label{eq:2.4}
&&\hat H=\hat H_\textrm{p}+\hat H_\textrm{i}\,,
\eea
where
\bea
\label{eq:2.5}
\hat H_\textrm{p} &=& NvU_{1 \textrm{seed}} - 2w_1 \sum\limits_{qq'}
\frac{\sigma_{q1}}{2} \frac{\sigma_{q2}}{2} \! \bigl( \delta_{{\bf R}_q{\bf R}_{q'}}\!\!
+ \!\delta_{{\bf R}_q + {\bf r},{\bf R}_{q'}} \bigr) \nonumber\\
&&+\frac12 \sum\limits_{{qq'},{ff'}}  J_{ff'}(qq') \frac{\langle \sigma_{qf}^{(n)}\rangle}{2}
\frac{\langle \sigma_{q'f'}^{(n)}\rangle}{2}
- \sum\limits_{{qq'},{ff'}}  J_{ff'}(qq') \frac{\langle \sigma_{q'f'}^{(n)}\rangle}{2}
\frac{\sigma_{qf}}{2}\nonumber\\
&&+\frac{1}{\sqrt{N}}\sum\limits_{qf}\sum\limits_{kl}\tau_{lf}(\vec{k})
\re^{\ri\vec{k}\vec{a_{q}}}\langle b_{kl}+b_{-kl}^{+}\rangle\frac{\sigma_{qf}^{n}}{2}\,,
\eea
\bea
\label{eq:2.6}\hat H_\textrm{i}&=& \sum\limits_{kl}\omega_{l}(\vec{k})b_{kl}^{+}b_{kl}
+\frac{1}{\sqrt{N}}\sum\limits_{qf}\sum\limits_{kl}\tau_{lf}(\vec{k})
\re^{\ri\vec{k}\vec{a_{q}}}\frac{\langle\sigma_{qf}\rangle}{2}\left(b_{kl}+b_{-kl}^{+}\right)\nonumber \\
&&-\sum\limits_{q}\sum\limits_{kl}\left[
\mu_{l}(\vec{k}){\rm cos}~ 5^{\circ}E_{1}^{\ast}+\mu_{l}(\vec{k}){\rm cos}~ 2^{\circ}E_s\right]\left(b_{kl}+
b_{-kl}^{+}\right)\frac{1}{\sqrt{N}}\re^{\ri\vec{k}\vec{a_{q}}}\,.
\eea

Using the Heisenberg equation of motion for the mean values of Bose-operators
\bea
\label{eq:2.7}
&& \frac{\rd}{\rd t}\langle b_{kl}^{\pm}\rangle = - \ri\langle [ b_{kl}^{\pm},\widehat{H}]\rangle,
\eea
we find that
\bea
\label{eq:2.8}
 \langle b_{kl}+b_{-kl}^{+}\rangle &=& -\frac{1}{\sqrt{N}}\sum\limits_{q'f'}\frac{2\tau_{lf}(-\vec{k})}{\omega_{l}(\vec{k})}
 \re^{-\ri\vec{k}\vec{a}_{q'}}\frac{\sigma_{q'f'}}{2}   \nonumber\\
&& +\sum\limits_{q'}\frac{2}{\omega_{l}(\vec{k})}\left[\mu_{l}(\vec{k}){\rm cos}~ 5^{\circ}E_{1}^{\ast}+\mu_{l}(\vec{k}){\rm cos}~ 2^{\circ}E_s\right]\frac{1}{\sqrt{N}}\re^{-\ri\vec{k}\vec{a}_{q'}}.
\eea

Taking into consideration the expression  (\ref{eq:2.5}), the Hamiltonian of system $\hat H_\textrm{p} $ takes the following form:
\bea
 \hat H_\textrm{p} &=& NvU_{1 \textrm{seed}} - 2w_1 \sum\limits_{qq'}  \frac{\sigma_{q1}}{2} \frac{\sigma_{q2}}{2}  \bigl( \delta_{{\bf R}_q{\bf R}_{q'}} + \delta_{{\bf R}_q + {\bf r},{\bf R}_{q'}} \bigr) \nonumber\\
&&+\frac12 \sum\limits_{{qq'},{ff'}}  \widetilde{J}_{ff'}(qq') \frac{\langle \sigma_{qf}\rangle}{2}\frac{\langle \sigma_{q'f'}\rangle}{2}
- \sum\limits_{{qq'},{ff'}} \widetilde{J}_{ff'}(qq') \frac{\langle \sigma_{q'f'}\rangle}{2}\frac{\sigma_{qf}}{2}
  \nonumber\\
&&-\sum\limits_{q}\left[
\mu{\rm cos}~ 5^{\circ}E_{1}^{\ast}+
\mu{\rm cos}~ 2^{\circ}E_{s}\right]\left( \frac{\sigma_{q1}}{2} + \frac{\sigma_{q2}}{2} \right).\nonumber
\eea
The following notations are used herein:
\bea
\label{eq:2.9}
\widetilde{J}_{ff'}(qq')&=& J_{ff'}(qq') +\frac{1}{{N}}\sum\limits_{kl}\frac{2\tau_{lf}(\vec{k})\tau_{lf'}(-\vec{k})}{\omega_{l}(\vec{k})}
\re^{\ri\vec{k}(\vec{a}_{q}-\vec{a}_{q'})},\\
\label{eq:2.10}\mu &=& \sum\limits_{q'}\sum\limits_{kl}\mu_{il}(\vec{k})
\frac{\tau_{lf}(\vec{k})}{\omega_{l}(\vec{k})}\frac{1}{\sqrt{N}}\re^{\ri\vec{k}(\vec{a_{q}}-\vec{a}_{q'})}.
\eea

Considering the symmetry of unary function of proton distribution
\[
\eta^{(1)x} = \langle \sigma_{q1} \rangle = \langle \sigma_{q2} \rangle
\]
and expanding the constant of long-range proton-proton interactions into a series
with respect to deformations  $\varepsilon_i$, $\varepsilon_4$, restricted to the linear summands:
\bea
\label{eq:2.11}
&& J_{ff'} = J + \sum\limits_{i=1}^3 \psi_{1i}\varepsilon_i + \psi_{4}\varepsilon_4\,,\\\nonumber
\eea
where  $J_{ff'} = \sum_{{\bf R}_q - {\bf R}_{q'}} J_{ff'}(qq')$ are Fourier transforms of
the  constants of long-range interaction, we get an output Hamiltonian   $\hat H$ in the following form:
\bea
\label{eq:2.12}
 \hat H &=& NvU_{1 \textrm{seed}} - 2w_1 \sum\limits_{qq'}  \frac{\sigma_{q1}}{2} \frac{\sigma_{q2}}{2}  \bigl( \delta_{{\bf R}_q{\bf R}_{q'}} + \delta_{{\bf R}_q + {\bf r},{\bf R}_{q'}} \bigr) - 2 N \nu_{1} (\eta^{(1)x})^2  - 2 \nu_{1} \eta^{(1)x} \left( \frac{\sigma_{q1}}{2} + \frac{\sigma_{q2}}{2}  \right)\nonumber\\
&& - \sum\limits_q \mu_{s}E_s \left( \frac{\sigma_{q1}}{2} + \frac{\sigma_{q2}}{2}  \right)-
\sum\limits_{q}\mu_{1}^{\ast}E_{1}^{\ast}\left( \frac{\sigma_{q1}}{2} + \frac{\sigma_{q2}}{2}  \right),
\eea
where the following notations are used:
\bea
&& \nu_{1} = \nu + \sum\limits_i \psi_{i}\varepsilon_i + \psi_{4}\varepsilon_4\,, \qquad \nu = \frac{J}{2}\,,\qquad \mu_{1}^{\ast}=
\mu{\rm cos}~ 5^{\circ}, \qquad \mu_{s}=\mu{\rm cos}~ 2^{\circ}.\nonumber
\eea

The approximation of two-particle cluster is used in order to calculate the physical
characteristics of PbHPO$_4$-type compound. In this approximation, the LHP thermodynamic
potential is as follows:
\bea
 G_1 &=& N U_{1\textrm{seed}} + 2 N \nu_{1} (\eta^{(1)x})^2  - k_\textrm{B} T \sum\limits_q \left\{\ln {\rm Sp}
 \re^{-\beta \hat H^{(2)}_{qA}} + \ln {\rm Sp} \re^{-\beta \hat H^{(2)}_{qB}} - \frac12 \ln {\rm Sp} \re^{-\beta \hat H^{(1)}_{q}}  \right\} \label{G2}\nonumber\\
&& -N v \sum\limits_i \sigma_i \varepsilon_i - Nv\sigma_4 \varepsilon_4\,,
\eea
where $\hat H^{(2)}_{q{A(B)}}$, $\hat H^{(1)}_{q}$ are two-particle and one-particle
Hamiltonians assigned by the following equations:
\bea
&& \hat H^{(2)}_{qA} = - 2w_1 \left( \frac{\sigma_{q1}}{2} \frac{\sigma_{q2}}{2}\right)
- \frac{x_{E}}{\beta} \left( \frac{\sigma_{q1}}{2} + \frac{\sigma_{q2}}{2}  \right), \label{H2}\\
&& \hat H^{(1)}_{q} = - \frac{\bar x_{E}}{\beta}\frac{\sigma_{qf}^{(1)}}{2}\,
. \label{H1}
\eea
The following notations are used herein:
\bea
\label{eq:2.16}
&& x_{E} = \beta \left( - \Delta^a + 2 \nu_{1} \eta^{(1)x}  + \mu_1^{\ast}E_1^{\ast} +\mu_{s}E_{s}\right),   \qquad \bar x_{E} = - \beta \Delta^a + x_{E}, \qquad \beta = 1/k_\textrm{B}T,
\eea
where $\Delta^a$ is the effective field formed dy the neighboring links beyond the  cluster borders.
Within the cluster approximation, the  $\Delta^a$ field is determined according to the condition of
self-consistency, i.e., the mean value of pseudospin   $\langle \sigma_{qf}^{(l)x} \rangle$ should
not depend on a particular Gibbs distribution (two-particle or one-particle Hamiltonian) according to which it is estimated:
\bea
&&\frac{{\rm Sp}\, \sigma_{qf}^{(l)} \re^{-\beta \hat H^{(2)}_{qA}}}{{\rm Sp} \, \re^{-\beta \hat H^{(2)}_{qA}}} =
\frac{{\rm Sp} \, \sigma_{qf}^{(l)} \re^{-\beta \hat H^{(1)}_{q}}}{{\rm Sp} \, \re^{-\beta \hat H^{(1)}_{q}}}\,. \label{Sp}
\eea

Then, based on (\ref{Sp}) and considering (\ref{H2}) and (\ref{H1}), we get an equation for the mean
value of pseudospin in the following form:
\be
\label{eq:2.18}
\eta^{(1)x} = \frac{{\sinh} x_{E}}{a_1 + {\cosh} x_{E}} = \frac{{\sinh} x_{E}}{D}\,,
\ee
where
\bea
&& x_{E }= \frac12 \ln \frac{1 + \eta^{(1)x}}{1 - \eta^{(1)x}} + \beta \nu_{1} \eta^{(1)x} + \frac{\beta\mu_1^{\ast}E_1^{\ast}}{2}+\frac{\beta\mu_{s}E_{s}}{2}\,,\qquad a_1 = \exp\left[{-\beta\left( w + \sum\limits_{i=1}^3\delta_i\varepsilon_i + \delta_4\varepsilon_4 \right)}\right]. \nonumber
\eea

\section{Static dielectric, piezoelectric, elastic and thermal characteristics \\ of  PbHPO$_4$ }

Having calculated the eigenvalues of two-particle and one-particle Hamiltonians, let us
present the thermodynamic potential (\ref{G2}) per unit cell in the following form:
\bea
 g_{1} &=& \frac{G_1}{N} = v U_{1 \textrm{seed}} - 2 k_\textrm{B}T \ln 2 -   (w + \delta_1 \varepsilon_1 + \delta_2 \varepsilon_2 + \delta_3 \varepsilon_3 + \delta_5 \varepsilon_5) \nonumber\\
&& +  \bigl( \nu + \psi_{1}\varepsilon_1 + \psi_{2}\varepsilon_2 + \psi_{3}\varepsilon_3 + \psi_{4}\varepsilon_4 \bigr) (\eta^{(1)x})^{2}  \nonumber\\
&& - \,k_\textrm{B}T \ln \left[ 1 - (\eta^{(1)x})^{2} \right] - 2k_\textrm{B}T\ln (a_1 + {\cosh} x_{E}) - v (\sigma_1\varepsilon_1 + \sigma_2\varepsilon_2 + \sigma_3\varepsilon_3) - v\sigma_4\varepsilon_4\,. \nonumber
\eea

Using the equilibrium equation
\[
\frac{1}{v} \left( \frac{\partial g_{1}}{\partial \varepsilon_i} \right)_{E_1} = 0, \qquad
\frac{1}{v} \left( \frac{\partial g_{1}}{\partial E_1} \right)_{\varepsilon_i} = -P_1\,, \qquad \frac{1}{v} \left( \frac{\partial g_{1}}{\partial E_1^{\ast}} \right)_{\varepsilon_i} = -P_1^{\ast}\,,
 \]
we get an equation for deformation $\varepsilon_i$, $\varepsilon_4$ and polarization $P_1$:
\bea
&&  0 = c_{i1}^{E0}\varepsilon_1 + c_{i2}^{E0}\varepsilon_2 + c_{i3}^{E0}\varepsilon_3 + c_{i4}^{E0}\varepsilon_4 - e_{1i}^0E_1 - \frac{\delta_i}{v} \frac{{\cosh} x-a_1}{a_1 + {\cosh} x} - \frac{\psi_{i}}{v} (\eta^{(1)x})^{2}, \label{sigma}\\
&& 0 = c_{14}^{E0}\varepsilon_1 + c_{24}^{E0}\varepsilon_2 + c_{34}^{E0}\varepsilon_3 +
c_{44}^{E0}\varepsilon_4 - e_{14}^0 E_1 - \frac{\delta_4}{v} \frac{ {\cosh} x -a_1}{a_1 + {\cosh} x} - \frac{\psi_{4}} {v}
(\eta^{(1)x})^{2}, \\
&& \label{eq:3.2}P_1^{\ast} = \chi_{11}^{\ast\varepsilon 0} E_1^{\ast} +  \frac{\mu_1^{\ast}}{v} \eta^{(1)x},\\\label{P2}
&& P_\textrm{s} =    \frac{\mu_{s}}{v} \eta^{(1)x}.
\eea

Based on the relation  (\ref{sigma}) and (\ref{P2}), we get the following thermodynamic
characteristics of LHP crystals, i.e., isothermal static susceptibility of a mechanically clamped crystal:
\bea
\label{eq:3.4}
&&  \chi_{11}^{^{\ast}\varepsilon T} = \left( \frac{\partial P_1^{\ast}}{\partial E_1^{\ast}} \right)_{\varepsilon_i} = \chi_{11}^{\varepsilon 0} + \frac{(\mu_1^\ast)^{2}}{2v} \frac{\beta \varkappa_{E}}{D - \varkappa_{E}\varphi^{\eta}}\,,
\eea
where the following notations are used:
\[
 \varkappa_{E} = {\cosh} x_{E} - \eta^{(1)x}{\sinh} x_{E}, \qquad \varphi^{\eta} = \frac{1}{1 - (\eta^{(1)x})^{2}} + \beta \nu_{1};
\]
isothermal coefficients of piezoelectric stress:
\bea
&&  e_{1i}^T = \left( \frac{\partial P_1}{\partial \varepsilon_i} \right)_{E_1} = e_{1i}^0 + \frac{\mu_1}{v} \frac{\beta (\psi_{i}\varkappa + \delta_i a_1)}{D - \varkappa\varphi^{\eta}} \eta^{(1)},\qquad e_{14}^T = \left( \frac{\partial P_1}{\partial \varepsilon_4} \right)_{E_1} = e_{14}^0 + \frac{\mu_1}{v} \frac{\beta (\psi_{4}\varkappa + \delta_4 a_1)}{D - \varkappa\varphi^{\eta}} \eta^{(1)};\nonumber
\eea
isothermal elastic constants at a constant field:
\begin{align*}
 c_{ij}^E &= \left( \frac{\partial \sigma_i}{\partial \varepsilon_i} \right)_{E_1} = c_{ij}^{E0} - \frac{\beta}{v} \delta_i\delta_j \frac{ a_1 {\cosh} x}{D^2} -  \frac{\beta}{v} \frac{ \eta^{(1)2}}{D - \varkappa \varphi^{\eta}}
\left[ \delta_i\delta_j \frac{\varphi^{\eta} a_1^2}{D} + \psi_{i}\psi_{j}\varkappa + \bigl( \delta_i\psi_{j} + \delta_j\psi_{i}\bigr) a_1 \right],  \nonumber\\
 c_{i4}^E &= c_{i5}^{E0} - \frac{\beta \delta_1\delta_4 {\cosh} x}{vD^2} - \frac{\beta \eta^{(1)2}}{v \bigl( D - \varkappa\varphi^{\eta}\bigr)}
\left[ \delta_i\delta_4 \frac{a_1^2\varphi^{\eta}}{D} + \psi_{i}\psi_{4}\varkappa +
\bigl( \delta_i\psi_{4} + \psi_{i}\delta_4\bigr) a_1 \right] , \nonumber\\
 c_{44}^{E0} &= c_{44}^{E0} - \frac{\beta \delta_4^2 {\cosh} x}{vD^2} - \frac{\beta \eta^{(1)2}}{v \bigl( D - \varkappa\varphi^{\eta}\bigr)}
\left[ \delta_4^2 \frac{a_1^2\varphi^{\eta}}{D} + \psi_{4}^2\varkappa +
2 \delta_4\psi_{4} a_1 \right]. \nonumber
\end{align*}

Other dielectric, piezoelectric and elastic characteristics of LHP may
be calculated using the above mentioned results. In particular, we may
get the matrix of isothermal compliance at a static field  $s_{ij}^E$,
which is inverse to the matrix of elastic constant  $c_{ij}^E$;
isothermal coefficients  of the  piezoelectric strain:
\be
\label{eq:3.5}
d_{1i}^T = \sum\limits_j s_{ij}^E e_{1i}^T\,, \qquad (i,j=1, 2, 3, 4),
\ee
isothermal dielectric susceptibility of free crystals:
\be
\label{eq:3.6}
\chi_{11}^{\sigma T} = \chi_{11}^{\varepsilon T} + \sum\limits_{i} e_{1i}^T d_{1i}^T\,.
\ee

In the LHP crystals, there takes phase a phase transition of the second order
from paraelectric phase into ferroelectric phase at temperature that satisfies the equation:
\be
\label{eq:3.7}
\exp\left(-\frac{w_1}{k_\textrm{B}T_\textrm{c}} \right)
= \frac{\nu_{1}}{k_\textrm{B}T_\textrm{c}} \,.
\ee
The molar entropy of a crystal conditioned by a proton subsystem is
obtained through a direct differentiation of thermodynamic potential:
\bea
&& S = - \frac{R}{2} \left( \frac{\rd g_{1}}{ \rd T} \right)_{\sigma_i} =  \frac{R}{2} \left\{ \ln2 + \ln \left[1 -
(\eta^{(1)})^2\right] + 2\ln D -
2\beta\nu_{1}(\eta^{(1)})^2 +
 \frac{2\beta w_{1}}{D}a_{1} \right\}, \label{S}
\eea
where $R$ is a universal gas constant.

The molar heat capacity of a hydrogen subsystem of LHP at a
constant strain is calculated by a direct differentiation of entropy (\ref{S}):
 \bea
 \label{eq:3.9}
 \Delta C_\textrm{p} &=&T\left( \frac{\partial S}{\partial T}\right)_{\sigma} \\ \nonumber
 &=&
 T\left(\frac{\partial S}{\partial T}\right)_{\eta,\varepsilon}+ T\left(\frac{\partial S}{\partial \eta}
\right)_{\varepsilon}\left(\frac{\rd\eta}{\rd T}\right)+\sum\limits_{i=1}^3  T\left(\frac{\partial S}{\partial\varepsilon_i}\right)_{\eta}  \left(\frac{\rd\varepsilon_i}{\rd T}\right) +T\left(\frac{\partial S}{\partial\varepsilon_4}\right)_{\eta} \left(\frac{\rd\varepsilon_4}{\rd T}\right).
 \eea

\section{Relaxational dynamics of PbHPO$_4$ crystal}

This section describes the dynamic phenomena in LHP at the application
of electrical field  $E_1^{\ast}$ to a crystal. While calculating the
dynamic characteristic of this kind of ferroelectrics we use a kinetic
equation  \cite{320x,321x, lev} based on the method of non-equilibrium
statistical Zubarev operator  \cite{38x}.

The kinetic equation for the mean values of pseudospin operator is as follows:
\be
\label{eq:4.1}
\frac{\rd}{\rd t} \langle \hat p_m \rangle = - \sum\limits_{qf} \sum\limits_{\mu\alpha} \left[ Q_{qf\mu\alpha}^{-}(\hat p_m) + {\tanh} \frac{\beta \Omega_{\mu}^{\alpha}}{2} Q_{qf\mu\alpha}^{+}(\hat p_m)\right] K_{\mu}^{\alpha}\,,
\ee
where
\bea
\label{eq:4.2}
&&Q_{qf\mu\alpha}^{\mp}(\hat p_m) = \langle \left\{ [ \hat p_m, \sigma_{qf}^{-\alpha} \bigl( \Omega_{\mu}^{\alpha'} \bigr) ], \sigma_{qf}^{\alpha} \bigl( \Omega_{\mu}^{\alpha} \bigr) \right\}^{\mp} \rangle_q\,, \\
&& K_{\mu}^{\alpha} = \int\limits_0^{\infty} dt e^{-\varepsilon t}\cos \Omega_{\mu}^{\alpha}t{\rm Re} \langle \bar u(t)\bar u^{+} \rangle_q\,,  \qquad \alpha=0,\pm 1,
\eea
while $\langle \bar u^{\alpha}(t)\bar u^{\alpha'} \rangle_q$ are
correlation function of a thermostat, $\sigma_{qf}^{\alpha} \bigl( \Omega_{\mu}^{\alpha} \bigr)$ is a Fourier component of the operator $\sigma_{qf}^{\alpha}(t)$, $\Omega_{\mu}^{\alpha}$ are eigenfrequencies of the Hamiltonian of pseudo-spin model (\ref{H2}), $\sigma_{qf}^0 = \sigma_{qf}$, $\sigma_{qf}^{\pm} = \sigma_{qf}^x \pm \ri\sigma_{qf}^y$.

Taking into account the time evolution law of pseudo-spin operators
$\sigma_{qf}^{\alpha}(t)$ and using the frequency presentation of these
operators, one may calculate the expressions for $Q_{qf\mu\alpha}^{\mp}\bigl( \hat p_{qf} \bigr)$
in another form. Finally, it allows us to rewrite the kinetic equation (\ref{eq:4.1}) as follows:
\bea
\label{eq:4.3}
&& \frac{\rd}{\rd t} \eta^{(1)x} = b_{11} \eta^{(1)x} + b_{12} \eta^{(2)x} + b_1\,, \nonumber\\
&& \frac{\rd}{\rd t} \eta^{(2)x} = b_{21} \eta^{(1)x} + b_{22} \eta^{(2)x} + b_2\,,
\eea
as well as the equation for a unary function in one-frequency approximation:
\be
\label{eq:4.4}
\frac{\rd}{\rd t} \eta^{(1)x} = -2K_0 \eta^{(1)x} + 2K_0 {\tanh} \frac{\beta \Omega_0}{2}\,.
\ee

The following notations are used herein:
\begin{align}
&b_{11} = - (K_1 + K_{-1}) + K_1Z_1 - K_{-1}Z_{-1}\,,&  &b_{12} = - K_1 + K_{-1}\,, & &b_1 = K_1Z_1 + K_{-1}Z_{-1}\,, \nonumber\\
&b_{21} = -2(K_1 - K_{-1}) + 2(K_1Z_1 + K_{-1}Z_{-1})\,, & &b_{22} = - 2(K_1 + K_{-1})\,, & &b_2 = 2(K_1Z_1 - K_{-1}Z_{-1})\,, \nonumber
\end{align}
\be
\label{eq:4.5}
K_{\mu} = K_{\mu}^{-1} + K_{\mu}^{1} = \int\limits_0^{\infty} \rd t \re^{-\varepsilon t} \cos \bigl( \Omega_{\mu} t \bigr) {\rm Re} \left\{ \langle \bar u^{-} (t) \bar u^{+}\rangle_q + \langle \bar u^{+} (t) \bar u^{-}\rangle_q \right\}, \qquad Z_{\mu} = {\tanh} \frac{\beta \Omega_{\mu}}{2}.\nonumber
\ee
In case $K_0 = K_{-1} = K_1 = {1}/{2\alpha}$, the system equations
obtained in this section are in agreement with the equations obtained
within the framework of a stochastic Glauber model  \cite{31}. Glauber
equations describe a physical situation at which the Fourier images of
the thermostat correlators are independent of the frequency  \cite{321x,320x}.

Thus, from equations  (\ref{eq:4.3}) and (\ref{eq:4.4}), we find that
\bea
&& \alpha \frac{\rd}{\rd t} \eta^{(1)x} = - (1 - P)\eta_{s}^{(1)x} + L, \label{deta2}\\
&& \alpha \frac{\rd}{\rd t} \eta^{(1)x} = - \eta^{(1)x} + {\tanh} \frac{\bar x}{2}\,, \nonumber
\eea
where the following notation is used:
\bea
\label{eq:4.7}
&&P = \frac12 \left[ {\tanh} \left( \frac{\beta w_1}{2} + \frac{x_{E}}{2} \right) - {\tanh}
\left(- \frac{\beta w_1}{2} + \frac{x_{E}}{2} \right) \right], \nonumber\\
&&L = \frac12 \left[ {\tanh} \left( \frac{\beta w_1}{2} + \frac{x_{E}}{2} \right) + {\tanh}
\left(- \frac{\beta w_1}{2} + \frac{x_{E}}{2} \right) \right].
\eea
Solving equations (\ref{deta2}) in the case of small deviations from the
equilibrium state, one may obtain the complex dielectric permittivity
of the hydrogen subsystem of LHP:
\be
\label{eq:4.8}
\varepsilon_{11}'^{\ast}(\omega) = \varepsilon_{11}^{\varepsilon 0} +
\frac{4\pi  \chi^{\ast}}{1 + (2\pi\nu\tau)^2}\,, \qquad
\varepsilon_{11}''^{\ast}(\omega) = \
\frac{4\pi \chi^{\ast}\pi\nu\tau}{1 + (2\pi\nu\tau)^2}
\ee
with the set of following notations
\begin{align}
\chi^{\ast} &= \frac{\mu_1^{\ast2}\beta}{2v} \left
\{\frac{2r\left[ 1 - P^{(0)}\right] - Y}{Yr} - \beta \left(\nu + \nu_{ab}\right)\right\}^{-1}\,\nonumber\\
\tau^{-1} &= \frac{1}{\alpha} \frac{Y r}{2r - Y} \left\{ \frac{2r
\left[ 1 - P^{(0)}\right] - Y}{Yr} - \beta (\nu + \nu_{ab}) \right\},\nonumber\\
 Y &= P^{(1)}  \eta^{(1)} + L^{(1)}, \qquad r = 1 - \bigl(  \eta^{(1)} \bigr)^2, \nonumber
\end{align}
\begin{align}
P^{(0)} &= \frac{1-a_1^2}{1+a_1^2 + 2a_1 {\cosh} {x}}\,,&
L^{(0)} &= \frac{2a_1 {\sinh} {x}}{1+a_1^2 + 2a_1 {\cosh} {x}}\,,\nonumber \\
P^{(1)} &= - \frac{4a_1(1-a1^2){\sinh} {x}}{[1+a_1^2 + 2a_1 {\cosh} {x}]^2}\,, &
L^{(1)} &= \frac{4a_1[2a_1 + (1+a_1^2) {\cosh}{x}]}{[1+a_1^2 + 2a_1 {\cosh} {x}]^2}\,. \nonumber
\end{align}

\section{ Comparison of numerical calculations with experimental data. \\
Discussion of the results obtained}

Prior to the discussion of the developed theory, it should be noted that this
theory, strictly speaking, holds for deuterated ferroelectric LDP.
Thermodynamic and dynamic characteristics of hydrogen-bonded ferroelectrics
taking tunneling $ \Omega$ into account, are essentially defined by an effective
parameter of tunneling  $\bar \Omega$, which is renormalized by short-range
interactions   \cite{147pok}. Here, $\bar \Omega \ll \Omega$, i.e., an essential
suppression of tunneling by short-range interactions takes place. Then, let us
assume that the theory proposed by us holds for LHP crystals as well taking into
consideration, in particular, the relaxational type of dispersion in LHP.

Unfortunately, the elastic constants of the LHP crystal have not been experimentally
determined so far. That is why it is impossible to specify the seed elastic constants
and hence to calculate, based on the proposed theory, the piezoelectric coefficients,
susceptibility of a mechanically free crystal, elastic constants in ferroelectric phase.
In order to perform the numerical calculations of temperature and frequency
dependencies of respective physical characteristics of LHP, the values of the following parameters should be specified:
\begin{itemize}
 \item parameter of a two-particle cluster $w$;
 \item parameter of long-range interaction $\nu$ ;
 \item effective dipole moment $\mu_{s}$;
 \item the seed dielectric susceptibilities $\chi_{11}^{ 0}$;
 \item parameter $\alpha$ that defines the time scale of relaxation processes.
\end{itemize}

In order to determine the above mentioned parameters, let us use temperature
dependencies of experimental physical characteristics, namely  $T_\textrm{c}$, $P_s(T)$ \cite{Negran},
$\varepsilon_{11}(0,T)$  \cite{Negran}.
The value of the effective dipole moment  $\mu_{s}$ is determined through the
agreement of the theory with the experiment for polarization of saturation.
In paraelectric phase, we determine  $\mu_{1+}^{\ast}$ by agreeing the theory
with the experiment for $\varepsilon_{11}^\sigma(T)$.

\begin{table}[!t]
\caption{A set of parameters of the theory for LHP and LDP crystals.\label{tab}}
\vspace{2ex}
\begin{tabular}{|c|c|c|c|c|c|c|c|c|c|c|c|c|}
\hline\hline
 &$T_\textrm{c}$ & ${w}/{k_\textrm{B}}$ & ${\nu}/{k_\textrm{B}}$ & $\mu_\textrm{s}$, $10^{-18}$ & $\mu_{1\textrm{p}}^{\ast}$, $10^{-18}$ & $\chi_{11}^{\varepsilon 0}$ &  $P_\textrm{s}$ & $R_\textrm{s}$& $P_\textrm{p}$  & $R_\textrm{p}$\\
  &  (K)  &  (K)  & (K)  & ($\text{esu}\cdot\text{cm}$) & ($\text{esu}\cdot\text{cm}$)&  & (s)  & (s/K)  & (s)&(s/K) \\
\hline\hline
 PbHPO$_4$& 310 & 850 & 19.98 & 1.00 & 1.46 & 0.716 &0.339 & -0.009 & 0.745 & 0.002 \\
\hline
 PbDPO$_4$& 452 & 1450 & 2.19 & 1.05 & 2.00 & 0.716 & & &  &  \\
\hline\hline
\end{tabular}
\end{table}

\begin{figure}[!t]
\centerline{
\includegraphics[scale=0.7]{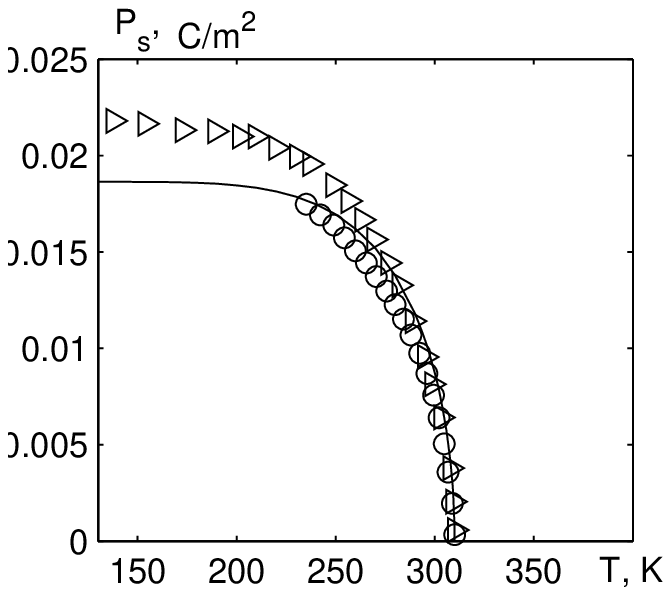}
\includegraphics[scale=0.7]{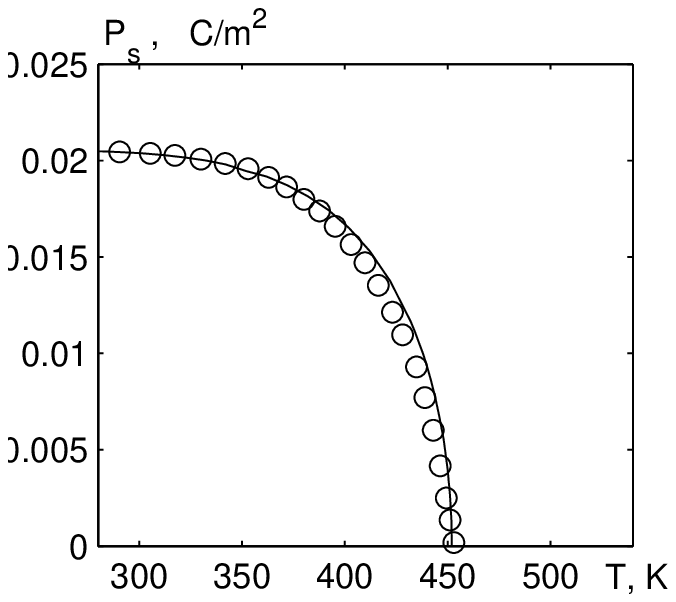}
}%
\caption{Temperature dependencies of spontaneous polarization of LHP~--- $\circ$ \cite{Negran}, $\diamond$  \cite{Smu} and LDP~--- $\circ$   \cite{Negran}.} \label{P_s}%
\end{figure}

\begin{figure}[!t]
\centerline{
\includegraphics[scale=0.7]{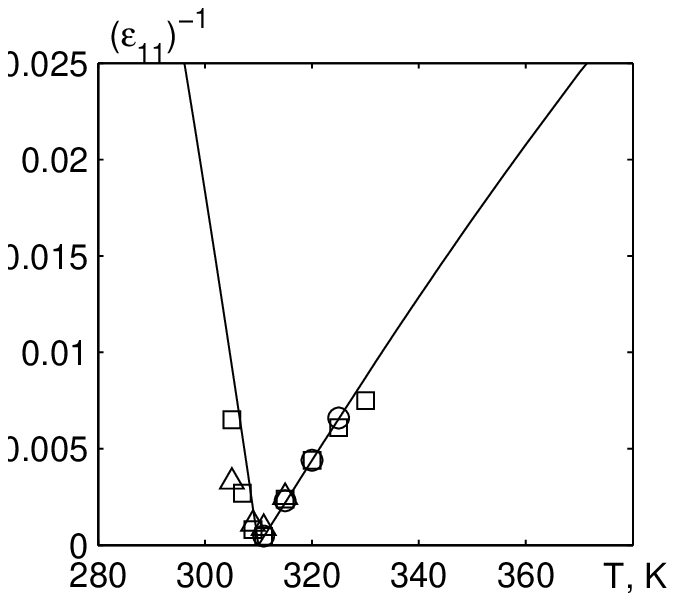}
\includegraphics[scale=0.7]{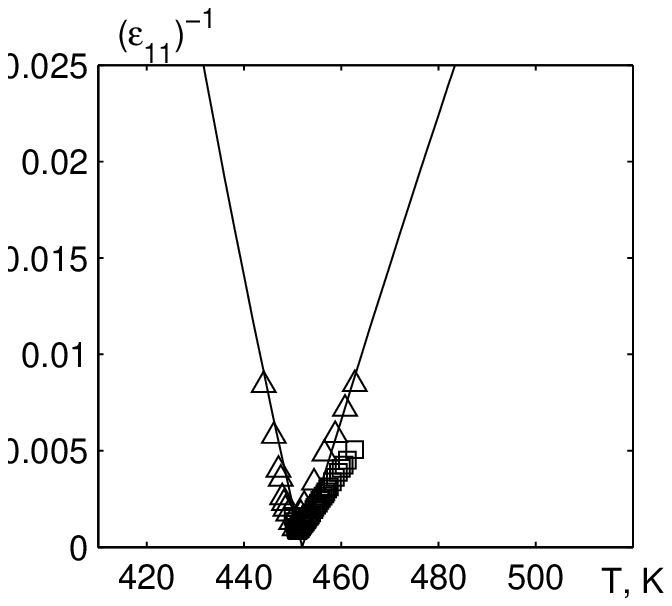}
}%
\caption{Temperature dependencies of  dielectric permittivities of LHP~--- $\circ$
 \cite{Negran}, $\square$  \cite{Deg}, $\vartriangle$  \cite{Nak} and LDP~--- $\vartriangle$
  \cite{Car}, $\square$  \cite{Deg}.} \label{eps11}%
\end{figure}

\begin{wrapfigure}{i}{0.5\textwidth}
\centerline{
\includegraphics[scale=0.75]{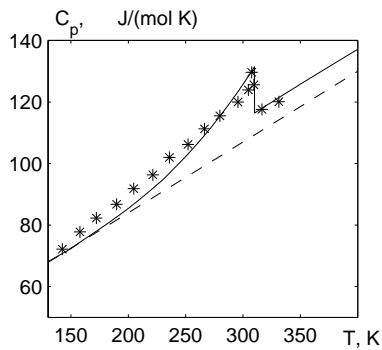}
}%
\caption{Temperature dependence of heat capacity for LHP crystal:  $\ast$,  \cite{Lop}.} \label{CpD}%
\end{wrapfigure}
Parameter $\alpha$ is defined from the condition that theoretically calculated
curves of $\varepsilon_{11}(\omega)$ should agree with the experimentally obtained curves.
In is assumed that parameter  $\alpha$ slightly changes with temperature:
\[
\alpha = [P + R(\Delta T)]\cdot 10^{-14}, \qquad \Delta T = T - T_\textrm{c}\,.
\]
The unit cell volume of LHP is taken to be equal to  $v = 0.1788\cdot 10^{-21}$~cm$^3$,  LDP~--- $v = 0.1800\cdot 10^{-21}$~cm$^3$.

The set of optimal parameters obtained this way is presented in table 1.

Let us now discuss the obtained results. Figure \ref{P_s} shows temperature
dependencies of spontaneous polarization of LHP and LDP crystals together with the experimental data.
It is seen in the figure that the data in papers    \cite{Negran} and  \cite{Smu}
disagree between themselves.
There is a good description of temperature dependencies of spontaneous polarization
obtained in paper  \cite{Negran}. Polarization of saturation increases at the
growth of the degree of deuteration  $x$.

\begin{figure}[!t]
\begin{center}
\includegraphics[scale=0.65]{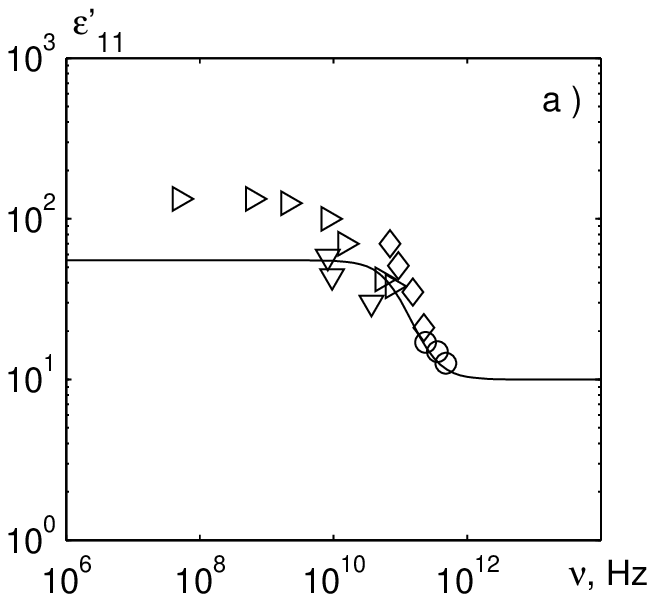}
\includegraphics[scale=0.65]{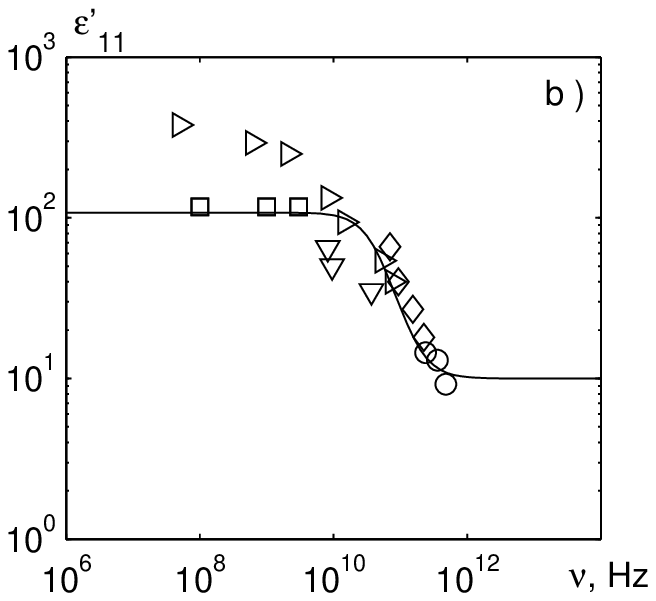}
\includegraphics[scale=0.65]{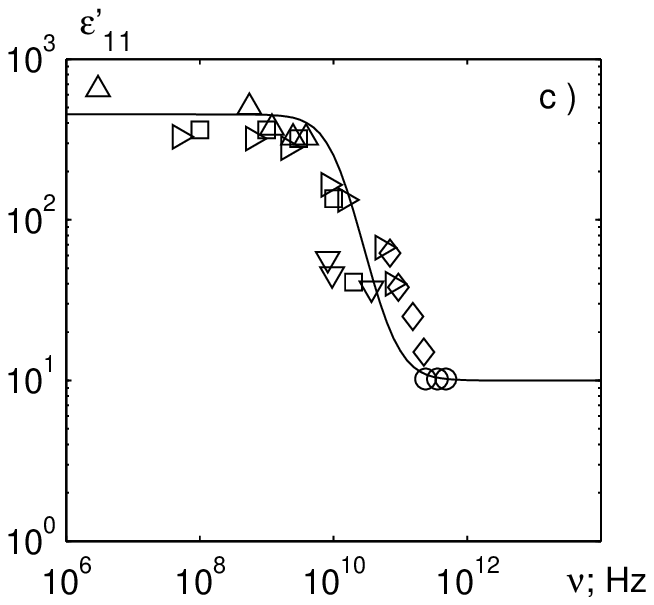}
\includegraphics[scale=0.65]{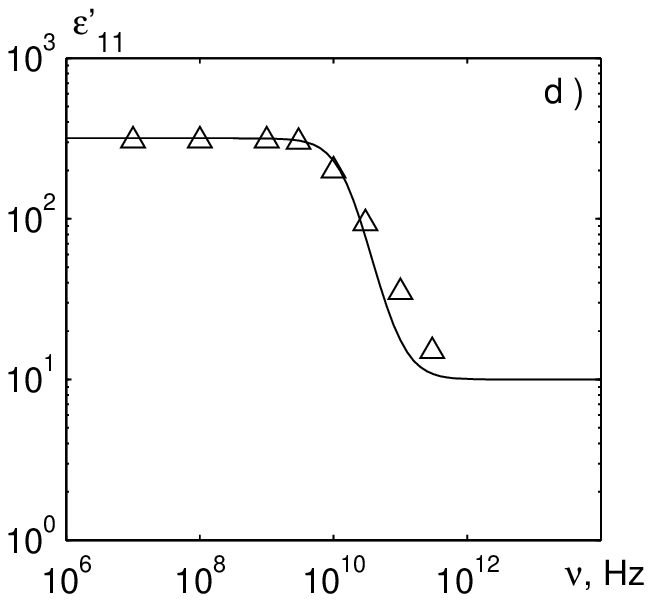}
\includegraphics[scale=0.65]{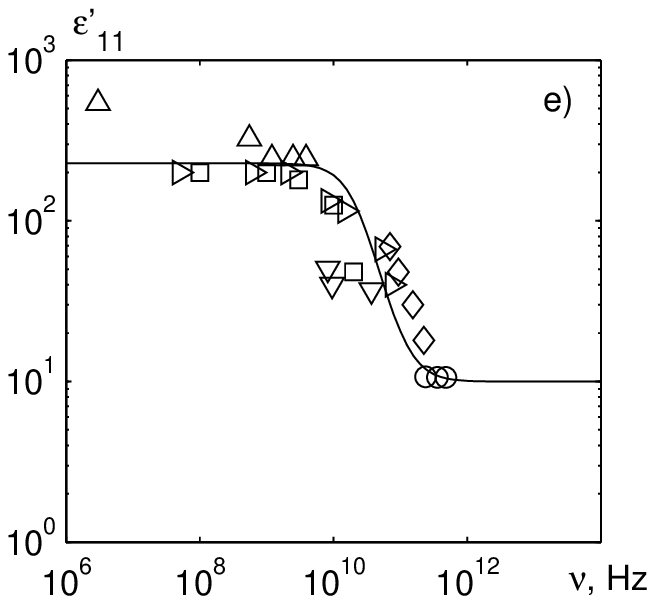}
\includegraphics[scale=0.65]{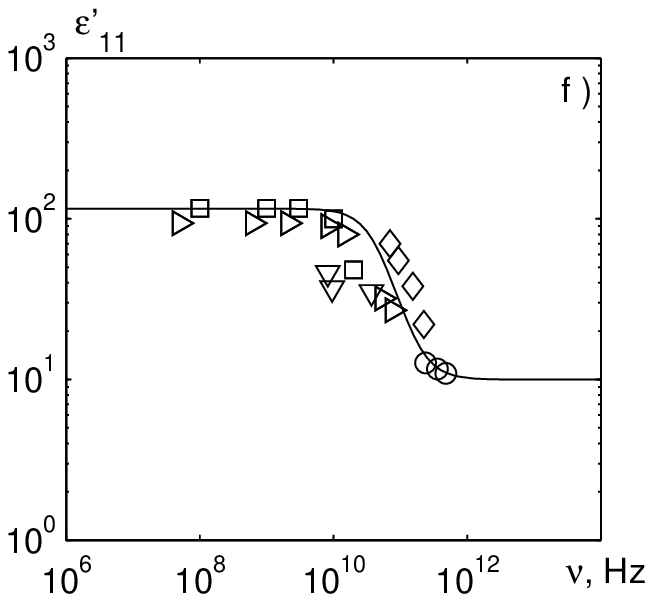}
\end{center}
\vspace{-1mm}
\caption[]{Frequency dependencies of real  $\varepsilon'_{11}$ part of dielectric
permittivity of LHP at $\Delta T$: a)  $-10$~K; b)  $-5$~K; c) 5~K; d) 7.16~K; e) 10~K; f)
 20~K and experimental data: $\Box$~---  \cite{Deg}, $\diamond$~---  \cite{Bri},
$\circ$~---  \cite{Kro}, $\bigtriangledown$~---  \cite{Koc}; $\triangleright$~---  \cite{Gri}, $\bigtriangleup$~---  \cite{Miz}.} \label{eps11re-10vci1}
\end{figure}

\begin{figure}[!b]
\begin{center}
\includegraphics[scale=0.65]{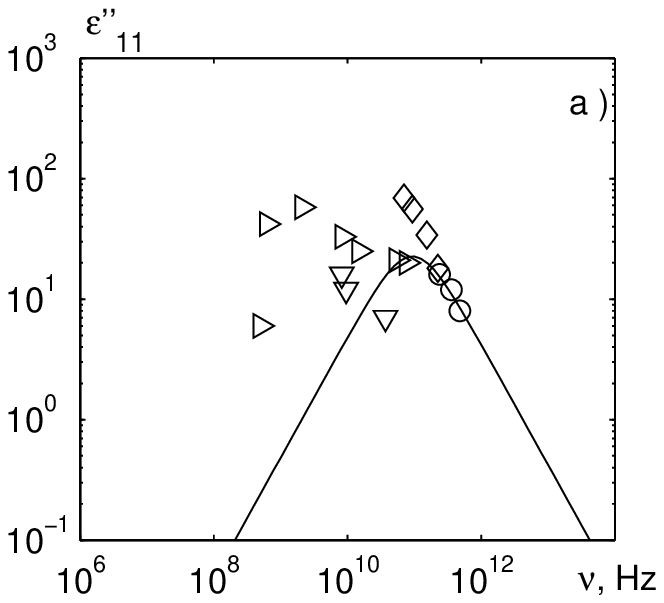}
\includegraphics[scale=0.65]{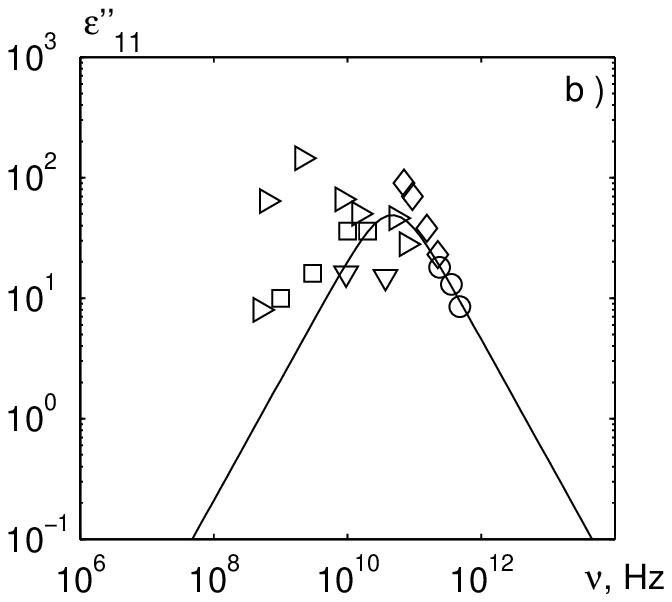}
\includegraphics[scale=0.65]{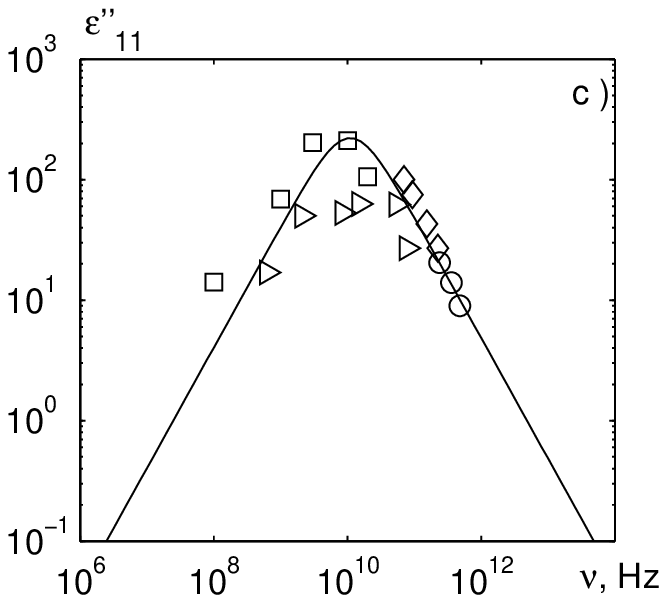}
\includegraphics[scale=0.65]{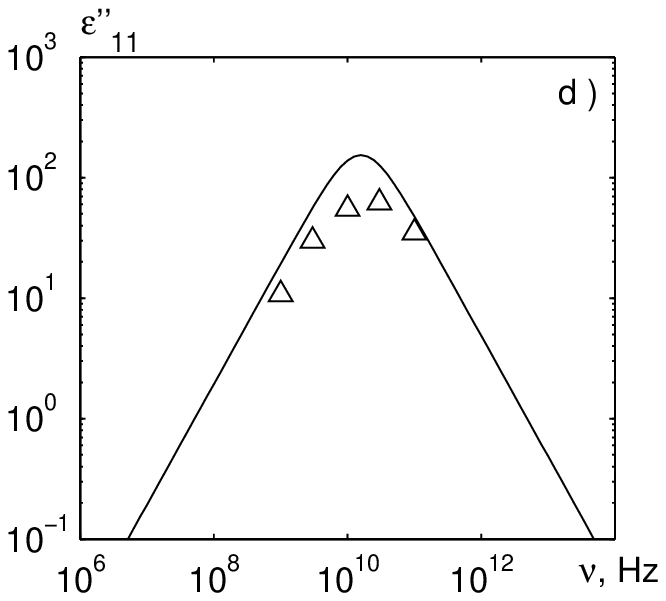}
\includegraphics[scale=0.65]{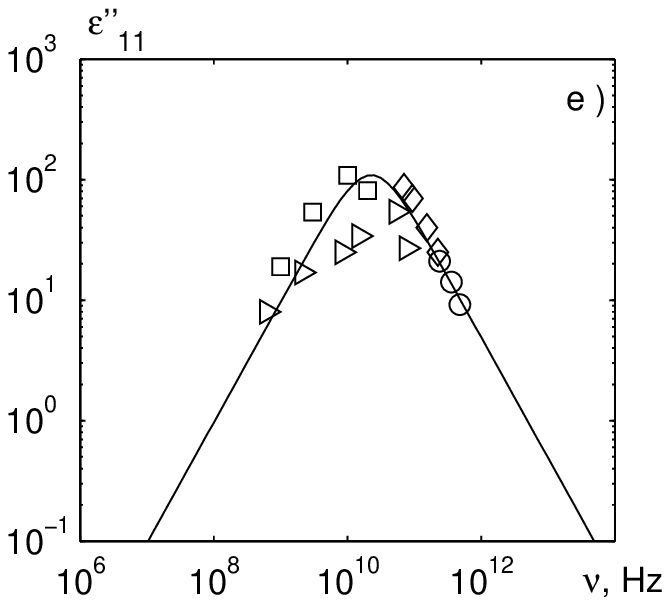}
\includegraphics[scale=0.65]{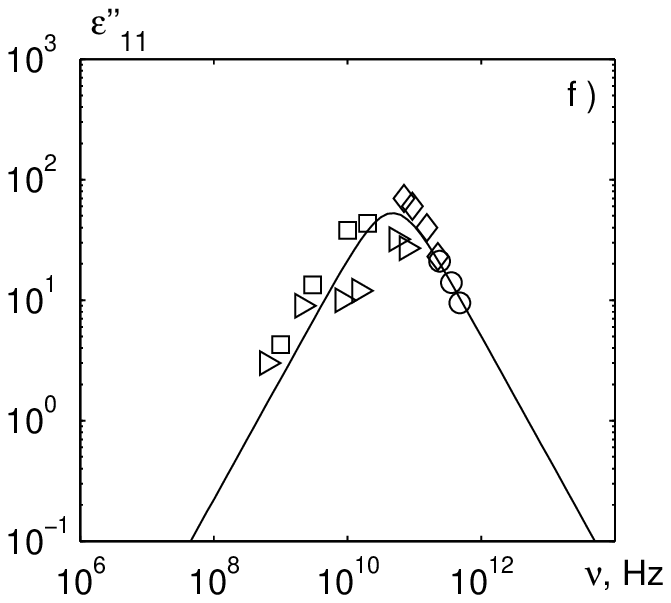}
\end{center}
\vspace{-1mm}
\caption[]{Frequency dependencies of imaginary  $\varepsilon''_{11}$ part of
dielectric permittivity of LHP at $\Delta T$: a) $-10$~K; b) $-5$~K; c) 5~K;
d) 7.16~K; e) 10~K; f) 20~K and experimental data: $\Box$~---  \cite{Deg},
$\diamond$~---  \cite{Bri};  $\circ$~---  \cite{Kro}; $\bigtriangledown$~---
 \cite{Koc}; $\triangleright$~---  \cite{Gri}, $\bigtriangleup$~---  \cite{Miz}.} \label{eps11im-10vci1}
\end{figure}

Figure  \ref{eps11} presents temperature dependencies of static dielectric
permittivities of mechanically clamped  $\varepsilon_{11}^{\ast}(0,T)$ crystals
of LHP and LDP calculated based on the microscopic theory, as well as the results
of experimental studies   \cite{Negran,Smu,Nak}.

As it is seen in figure \ref{eps11}, the results of theoretical calculations
of $\varepsilon_{11}^{\ast}(0,T)$ on the whole show good quantitative agreement
with experimental data obtained by the  authors  \cite{Negran,Deg,Nak}.

The temperature dependence of heat capacity for LHP crystal, together with
experimental data of paper  \cite{Lop} are presented in figure \ref{CpD}.

The dashed line shows the effective  lattice contribution  $C_0$ into the
heat capacity which is estimated by us as the average of the difference
$C_\textrm{p}(T) - \Delta C_\textrm{p}(T)$. Based on the proposed model and using the
theory parameters (table~\ref{tab}), a quantitatively good description of the data
of paper  \cite{Lop} has been achieved. The amount of the calculated heat
capacity jump well correlates with the experiment.

 The measured temperature dependencies of a real and imaginary parts of dielectric
 permittivity at different frequencies for LHP are presented in papers \cite{Kro,Bri,Koc,Gri,Miz,Deg,Sap}.

Figures  \ref{eps11re-10vci1} and \ref{eps11im-10vci1}   present the results
of calculations of frequency dependencies  $\varepsilon'_{11}(\nu)$ and $\varepsilon''_{11}(\nu)$,
respectively, for LHP  as well as experimental data.

In these figures it is seen that there is a significant data scattering
obtained in the experimental papers  \cite{Kro,Bri,Koc,Gri,Miz,Deg}. The best
agreement for our calculated data of $\varepsilon'_{11}(\nu)$ and $\varepsilon''_{11} (\nu)$
and the experiment was reached with the results published in papers  \cite{Deg,Bri,Kro}.

\begin{figure}[!t]
\begin{center}
\includegraphics[scale=0.7]{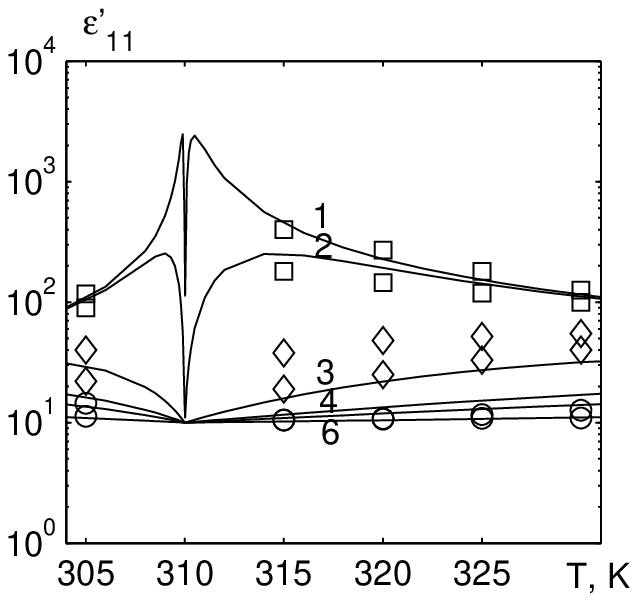}
\includegraphics[scale=0.7]{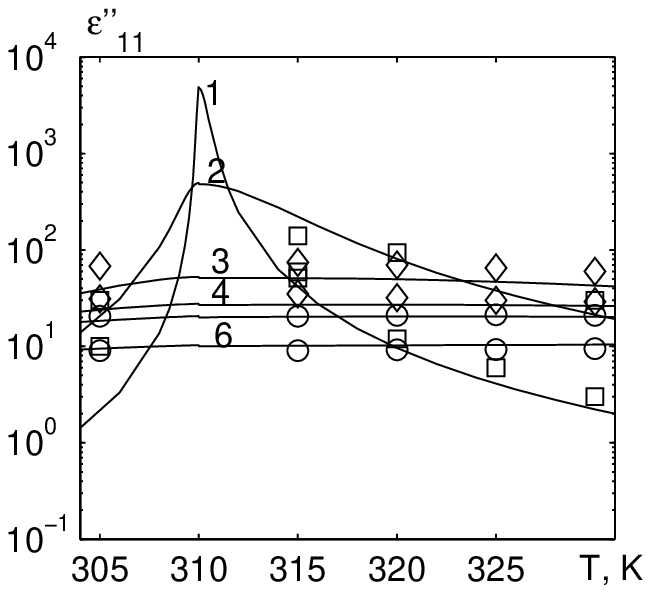}
\end{center}
\caption[]{Temperature dependencies of real  $\varepsilon_{11}'(\nu,T)$ and
imaginary  $\varepsilon_{11}''(\nu,T)$ parts of  dielectric permittivity of
LHP crystal  at different frequencies $\nu$ (GHz): 1~--- 1, $\square$ \cite{Deg};
10~--- 2, $\square$ \cite{Deg}; 94~--- 3; $\lozenge$  \cite{Bri}; 179~--- 4, $\lozenge$
 \cite{Bri}; 240~--- 5, $\circ$ \cite{Kro}; 480~--- 6, $\circ$ \cite{Kro}.} \label{eps11re_T}
\end{figure}

Figure  \ref{eps11re_T} presents temperature dependencies of real
$\varepsilon_{11}'(\nu,T)$ and  imaginary  $\varepsilon_{11}''(\nu,T)$
parts of dynamic dielectric permittivity at different frequencies of LHP
crystal as well as the experimental data.


As seen in this figure, there is quite acceptable agreement between our theoretical
results and the majority of experimental data. A significant disagreement is observed
only with the experimental data of Briskot et al.  \cite{Bri}. The reason is a rather poor
experimental accuracy (order of 40--50~\%) of their experimental setup.

\section{Conclusions}
The present paper, based on a modified model of proton ordering that does not take
into consideration the proton tunneling on hydrogen bonds in the approximation of a
two-particle cluster, describes the theory of thermodynamic and dielectric, piezoelectric,
elastic and dynamic properties of one-dimentional ferroelectrics  of PbHPO$_4$-type.
Optimal sets of model parameters have been found that make it possible to describe
the available corresponding experimental data for LHP and LDP crystals.


\vspace{-5mm}

\ukrainianpart
\title{ Термодинамічні  та динамічні діелектричні властивості одновимірних  сегнетоелектриків з водневими зв'язками типу PbHPO$_4$}
\author{І.Р. Зачек\refaddr{label2}, Р.Р. Левицький\refaddr{label1}, Я.Й. Щур\refaddr{label1}, О.Б. Біленька\refaddr{label2}}
\addresses{
\addr{label1} Інститут фізики конденсованих систем НАН України, вул.
І~Свєнціцького, 1,  79011 Львів, Україна
\addr{label2} Національний
університет ``Львівська політехніка'', вул.~С. Бандери, 12, 79013 Львів, Україна }

\makeukrtitle

\begin{abstract}
\tolerance=3000%
У рамках модифікованої моделі протонного впорядкування з врахуванням взаємодії протонів з нормальними коливаннями гратки одновимірних сегнетоелектриків з водневими зв'язками типу PbHPO$_4$ з врахуванням лінійних за деформаціями кристалу $\varepsilon_i$ і $\varepsilon_4$ внесків в енергію протонної системи, але без врахування тунелювання в наближенні двочастинкового кластера розраховано і досліджено їх термодинамічні і динамічні характеристики. Отримано добрий кількісний опис температурної залежності поляризації, статичної діелектричної проникності кристалів PbHPO$_4$ та PbHDO$_4$, теплоємності і частотної залежності динамічної діелектричної проникності при різних температурах  кристалу  PbHPO$_4$.
\keywords сегнетоелектрик, діелектрична проникність, п'єзоелектричний коефіцієнт, PbHPO$_4$
\end{abstract}

\end{document}